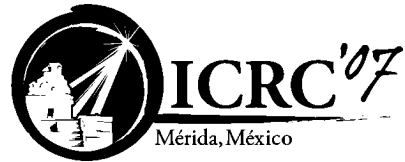

# A study of the Forbush decrease event of September 11, 2005 with GRAND

J. POIRIER, M. HERRERA, P. HEMPHILL, C. D'ANDREA
*Physics Department, University of Notre Dame, Notre Dame, IN 46556 USA*
*poirier@nd.edu*

**Abstract:** Project GRAND, a proportional wire chamber array, is used to examine the decreased counting rate of ground level muons during the Forbush decrease event of September 11, 2005. Data are presented and compared to that of other cosmic ray monitors. A directional study of the Forbush decrease was undertaken and precursor anisotropies to this geomagnetic storm were studied utilizing GRAND's angular resolution.

## Introduction

A Forbush decrease (FD) is an intensity depression in cosmic ray count caused by a coronal mass ejection (CME) [2]. When a CME arrives at the Earth, a sudden storm commencement (SSC), an intensification in the low latitude ground-based magnetic field intensity [9], can be detected and used as the onset time of a geomagnetic storm [4]. On September 11, 2005, the National Geophysical Data Center [7] website reported an SSC at 1:14 UT. The event was ranked by 12 different magnetic observatories as "very remarkable" [7]. In this paper, data are analyzed from Project GRAND for evidence of a FD event associated with this CME/SSC event. These data are compared to data taken from the Nagoya Multi-directional Telescope and the Oulu Neutron Monitor. A directional study of the FD event was undertaken utilizing the angular resolution capabilities of GRAND in order to study precursor anisotropies to SSCs which can result from a "loss cone" effect, during which the detector observes the deficiency of particles which trace to the cosmic-ray-depleted region behind the shock [4].

## Experiment

Project GRAND is an array of 64 proportional wire stations located adjacent to the University of Notre Dame campus at $41.7^{o}$ N and $86.2^{o}$ W at an altitude of 220 m above sea level. Each station contains four PWC plane pairs. The array is actually two experiments: the tracking of low energy single muon events, and the recording of high energy extensive air showers. Only the single muon data are considered for this study. Each station contains eight 1.29 $m^2$ proportional wire chamber (PWC) planes yielding a total active area of 82 $m^2$. Each pair contains a horizontal plane of wires running north-south, and another plane of wires running east-west. When a charged particle (such as a secondary muon created by cosmic rays) passes through the chamber, it leaves a trail of ions. These ions are then accelerated toward the closest signal wire; as they are accelerated, they collide with more gas molecules and release more charge in a process known as gas amplification which further increases the charge collected on the signal wire. A small current is formed on this signal wire which is amplified. This electronic signal denotes the position of the charged particle which passed near this wire. By comparing hit wires in planes vertically above each other, the angle of the muon track can be reconstructed to within 0.26°, on average, in each of two projected planes: up/east and up/north. A 50 mm thick steel plate is situated above the bottom two PWC planes allowing for the distinction of muon tracks from the electron tracks which stop, shower, or are deflected by the steel. The array collects data at a rate of ~2000 identified muons per second. Added details are available at: http://www.nd.edu/~grand.

## Data and observations

GRAND data from September 9, 2005 at 5.0 UT through September 20, 2005 at 5.0 UT were considered for this study. The muon count for this time interval was examined in one hour bins. In order to ensure experimental accuracy, the r.m.s. deviation was calculated for each station and compared to its expected statistical variation. A histogram of these ratios was compiled for each station for each day, and a cutoff was determined such that if its ratio was higher than this cutoff it was excluded in the analysis. There were 30 stations left in the analysis after this severe cut, ensuring that time changes in the sum-of-huts muon count rate could only be the result of actual muon rate changes.

The muon counts were pressure corrected using data from the National Climactic Data Center [6]. The pressure corrected muon data from Project GRAND are shown in Figure 1a with the points plotted at the end of each hour.

A drop of approximately 3% is demonstrated by the GRAND data on September 11, 2005. The greatest decrease in count in one hour (the most negative slope of the plot) occurs from 3.5 to 4.5 UT on September 11, 2005. There are no points near September 9, 2005 23:00 UT and September 20, 2:00 UT due to data interruptions.

Other cosmic ray monitors also observed this FD. Data from the Nagoya Multidirectional Muon Telescope [5] and the Oulu Neutron Monitor [8] are compared in Figures 1b and 1c (here the points are plotted at the beginning of each hour). The Nagoya Multidirectional Muon Telescope detected an approximately 5% drop on September 11, 2005 with the greatest decrease in count in one hour occurring between 4.5 UT and 5.5 UT. The Oulu data demonstrate a decrease of approximately 11% on September 11, 2005, with the greatest decrease in count in one hour occurring between 1.5 UT and 2.5 UT. Because Oulu measures secondary neutrons as opposed to muons, its primary energy sensitivity is influenced by different mechanisms producing neutrons rather than muons. In addition, Oulu has a lower geomagnetic cutoff rigidity (0.8 GV) compared to Nagoya (11.5 GV) and GRAND (1.9 GV) As a result of these factors, Oulu is more sensitive to lower energy primaries than muon stations and therefore can account for its larger drop in count rate; the percentage drops of Nagoya and GRAND are similar.

Utilizing GRAND's angular resolution, the directions of the secondary muon tracks were studied. The sky was divided into a three-by-three grid of viewing directions. The grid is defined by two orthogonal projected angles: the angle from zenith going north, and the angle from zenith going east. Divisions between the viewing directions were made at 9.7° and 61° [3] in both the north and east directions such that each viewing direction bin would have roughly the same number of muon counts per hour during a background day (and hence similar statistical precision). Directional data from ten background days were taken as a baseline of comparison. The percentage of counts coming from each viewing direction for each hour was calculated for background days and the day in study. A mean percentage was calculated for each viewing direction for each hour, and then compared to hours just before and during the FD event. The results of the study are displayed in Figure 3.

The rate shows an intensification in counts in the SW corner of the sky. Angular deviations were also observed when a nine-day base was compared to adjacent days that were supposedly quiet in terms of their angular distributions. This, as well as other unsuccessful attempts to find "quiet days", leads one to conclude that, to the level of statistical precision available to GRAND, there are more days of statistically significant angular activity than expected.

A possible precursor to a geomagnetic storm is the "loss cone anisotropy" which would occur hours before the sudden storm commencement [4]. A "loss cone" is a deficiency in counts from directions that make small angles relative to the interplanetary magnetic field (IMF). In an attempt to observe the loss cone anisotropy, the pitch angles of the GRAND viewing directions were calculated relative to the IMF as measured by the Advanced Composition Explorer, accessed through the Omniweb Interface [1, 3]. Again, a ten-day background percentage was formed for each viewing direction in each hour. These average percents were compared to the hourly percentages for each viewing direction during September 11, 2005. The deviations were then represented on a bubble plot with pitch angle vs. time, with the size of each bubble representing the amount of deviation from the baseline average. This bubble plot is displayed in Figure 2.

The overall shape of the data comes from the fact that not all pitch angles are visible to GRAND at a given time. Unfortunately, GRAND was not at the right latitude to measure pitch angles close to 0° for several hours prior to the September 11, 2005 FD event.

## Conclusions

Project GRAND sees a Forbush decrease event when examining the secondary muon counting rate at ground level between September 9, 2005 at 5.0 UT and September 20, 2005 at 5.0 UT. Similar decreases are recorded by the Nagoya Multi-directional Muon Telescope and the Oulu Neutron Monitor. The differences in the magnitude of the drop between the muon and neutron detectors come from the different mechanisms that produce muons and neutrons, as well as the differences in geomagnetic cutoff rigidity. A directional study of the FD event discovers an increase coming from the SW section of the sky. There were, however, unexpected variations in the angular distribution of muons during supposedly quiet days. This, coupled with other instances of unexpected variations during seemingly normal days leads one to conclude that quiet days, if they exist to the level of statistical precision available to GRAND, are less numerous than expected. Finally, to study loss cone anisotropy, a bubble plot which displays percent deviations as a function of pitch angle and time was created; however, GRAND was unable to observe the relevant pitch angles before the sudden storm commencement. The open spaces in Figure 2 are regions that are unobservable to GRAND at those times. This shows the need of a worldwide network of muon detectors. Future goals entail the inclusion of Project GRAND into such a worldwide muon detector network to detect geomagnetic storm precursors and more uniformly monitor the sun.

The authors wish to thank K. Munakata for the calculation of GRAND's asymptotic viewing directions. Thanks to the Oulu Neutron Monitor and the Sodankyla Geophysical Observatory, the National Climatic Data Center, the Advanced Composition Explorer, the National Geophysical Data Center, the Cosmic Ray Section, Solar-Terrestrial Environment Laboratory, and Nagoya University for the use of their data. The OMNI data were obtained from the GSFC/SPDF OMNI-Web interface at: http://omniweb.gsfc.nasa.gov. Thanks to the National Science Foundation for their support of the REU program. Project GRAND is currently funded through the University of Notre Dame and private donations.

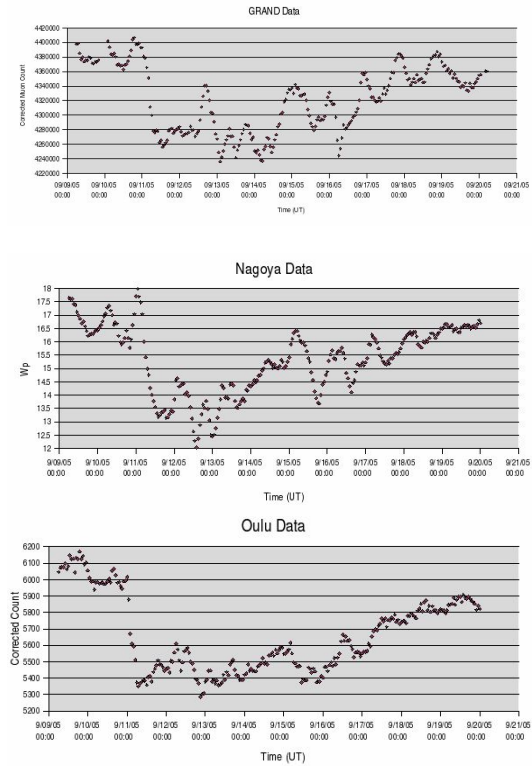

Figure 1: Count rate vs. time during the Forbush Decrease. a) top: GRAND muons; b) middle: Nagoya muons; c) bottom: Oulu neutron monitor. (For a) points plotted at the end of each hour; for b) and c) they are at beginning of each hour).

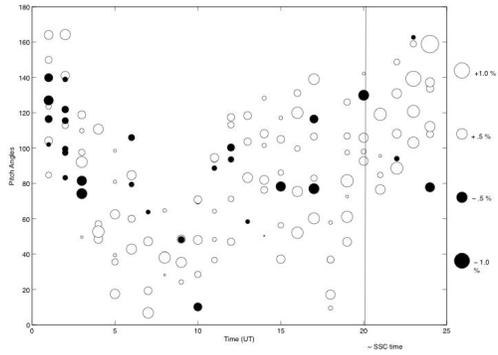

Figure 2: A bubble plot showing the angular deviation as a function of time; black circles are negative deviations; white circles are positive deviations. The scale for circle sizes is on the right side.

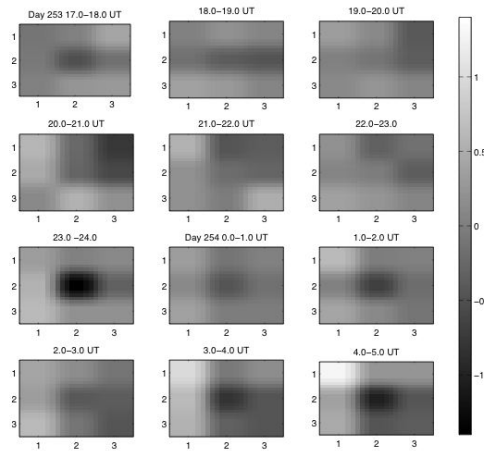

Figure 3: The percent deviations of the muon count for the viewing directions in three northward (down) channels and three eastward channels (right) for each hour around the FD.